\documentclass[conference]{IEEEtran}
\usepackage[small,bf]{caption}
\usepackage{color}
\usepackage{graphicx}
\usepackage{amsmath}
\usepackage{amssymb}
\usepackage{amsmath}
\usepackage{rotating}
\usepackage{verbatim}
\usepackage{fancyvrb}

\begin{document}

\title{Using Alloy to Formally Model and Reason About an OpenFlow Network Switch}

\author{
%\begin{center}
\small
\begin{tabular}[t]{ccccccccc}
{Saber Mirzaei}&&{Sanaz Bahargam}&&{Richard Skowyra}&&{Assaf Kfoury}&&{Azer Bestavros}\\[0.5 mm]
\tt{smirzaei@bu.edu} && \tt{bahargam@bu.edu}&& \tt{rskowyra@bu.edu} && \tt{kfoury@bu.edu} && \tt{best@bu.edu}\\~\\
\end{tabular} \\
Computer Science Department\\[1mm]
Boston University \\[2mm]
July 2013\\[4mm]
}

\maketitle
%\pagenumbering{arabic}

\begin{abstract}
Openflow provides a standard interface for partitioning a network into a data plane and a programmatic control plane.
While providing easy network reconfiguration, Openflow introduces the potential for programming bugs, causing network deficiency.
To study the behavior of OpenFlow switchs, we used Alloy to create a
software abstraction, describing the internal state of a network and its OpenFlow
switches. Hence, this work is an attempt to model the static and dynamic behaviour of
networks configured using OpenFlow switches.

\end{abstract}
\IEEEpeerreviewmaketitle
\section{Introduction}
\emph{Software-defined networking} (SDN) is a technique
used in computer networking to decouple the control plane and the data plane
of the network.
SDN employs a centralized software program to automatically configure
and control network switches. %without manually configuring individual
SDN helps researchers to write high level programs to control network behavior
instead of manually configuring and manipulating policies as in traditional network
switches.
OpenFlow is the first standard communication interface defined between the
control and forwarding layers of an SDN architecture.
OpenFlow is an open API to remotely control forwarding tables of switches by
adding and removing flow entries \cite{mckeown2008openflow}.
OpenFlow provides an easy interface for changing network configurations, but also enables the potential of introducing software bugs impacting network behavior.
This raises a number of questions: ``Does changing configurations via OpenFlow
cause any security breaches or inconsistencies in switches?
Can it cause undesired network behaviors unknown to the network operator?"

In this report, we will use Alloy, a lightweight modeling language, to create a
software abstraction describing the static structure and dynamic behavior of the
OpenFlow network switches.
Using this model, in future work we aim to answer the aforementioned questions about
OpenFlow.
The remainder of the paper is laid our as it follows.
		
In Section \ref{sec:background} we describe SDN in more
detail, and introduce the fundamental concepts of the Alloy language.

Section \ref{sec:switchStraucture} discusses the key components of an OpenFlow switch.
We explain the static model of an OpenFlow switch in Section \ref{sec:static} and the
dynamic model in Section\ref{sec:dynamic}.
Related work on SDN design and verification is reviewed in Section
\ref{sec:related}.
Finally, Section \ref{sec:conclusionFuture} concludes the paper and discusses our
future work.
\begin{comment}
\subsection{Motivation}
In traditional network, to change a policy or rule, network operator had to
manually change configurations in switches. Not only this process was time
consuming, but also changing configuration mistakenly was a common issue. Other
drawbacks include:
\begin{itemize}
\item Difficult to perform real world experiments on large scale production
networks
\item Rate of innovation in networks is slower/ lack of high level abstraction
\item Innovation is limited to vendor/ vendor partners
\item Huge barriers for new ideas in networking
\end{itemize}
\end{comment}

\section{Background}
\label{sec:background}
\subsection{Software Defined Networks}
In SDN the data forwarding plane is separated from control plane,
which is managed by a network OS.
The network OS (such as Nox \cite{NOX:SIGCOMM2008}, POX
\cite{POXController},Beacon \cite{beacon}, or ONIX \cite{ONIX:2010})
controls the whole network from a central point.
It controls the data plane via interfaces such as OpenFlow
\cite{mckeown2008openflow}.
Accordingly, the functionality of a network can be defined and changed after SDN
has been physically deployed.
Hence, changing network switch's rules, prioritizing, de-prioritizing or even
blocking/re-routing packet flows can be facilitated in a very
fine-grained level of control.

A software defined controller allows us to trace and manage specific flows in a flexible
approach based on packets' header information (such as packet's
source/destination address).

\begin{figure}[!]
\begin{center}
\includegraphics[scale=.7]%
    {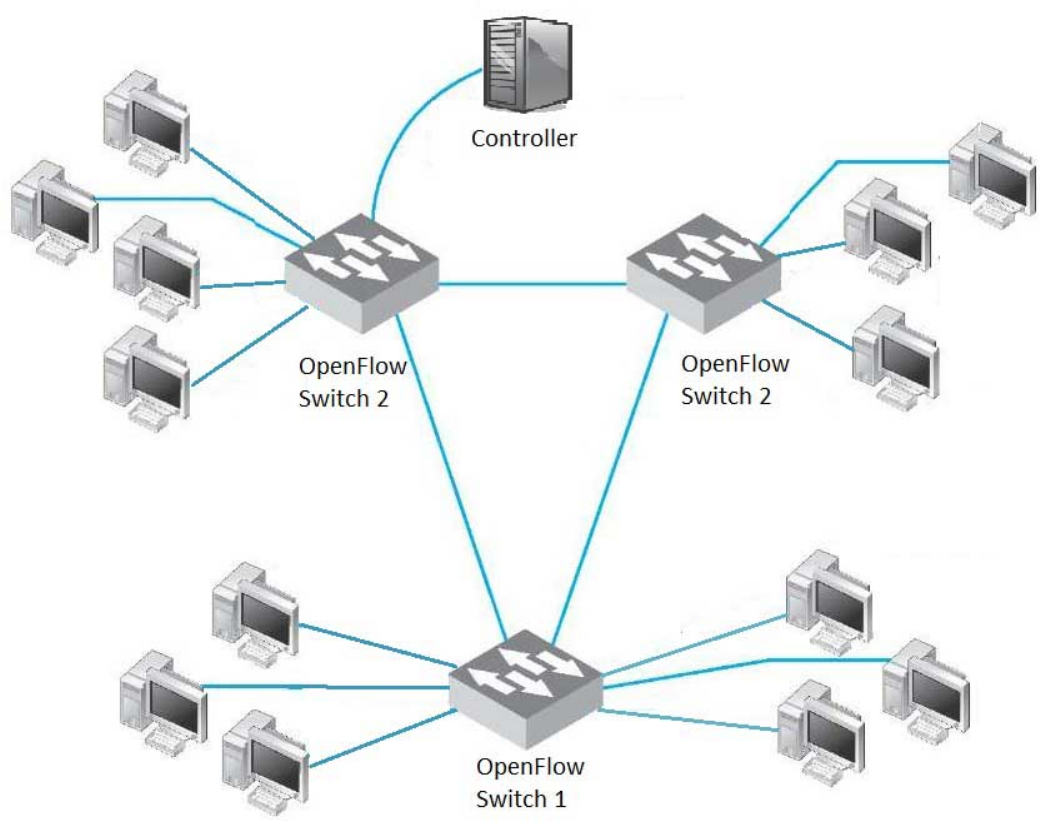}

\caption{Example of a network with three switches and one controller.}
\label{fig:openflowNetwork}
\end{center}
\end{figure}

OpenFlow is the first standard communications interface defined between the
control and forwarding layers of an SDN architecture.
OpenFlow facilitates the software defined routing of packets through the network
of switches and also provides sophisticated traffic management.
The basic idea of OpenFlow is to exploit the concept of flow-tables (already
used in Ethernet switches) for different applications such as implementing
firewalls, QoS and NAT.
Employing this notion, OpenFlow provides a protocol in order to program the
flow-tables for routing packets and managing flow traffics.
More importantly, using OpenFlow, network administrators can separate
production and research traffic.
Hence this gives the researcher the ability to implement and test new routing
protocols, security models or even alternatives to IP
\cite{mckeown2008openflow} on real-world networks.
Figure \ref{fig:openflowNetwork} shows a network of OpenFlow switches. In this
example, all the switches are managed by only one controller.
In general, based on OpenFlow specification, a switch can be controlled by more
than one Controller.\\
Every OpenFlow switch consists of at least three parts \cite{specv1}:
\begin{itemize}
  \item A set of Flow Tables (at least one table). Each table has a set of flow
entries. Each entry comprises a set of actions that will be applied to the
packet when it matches that entry.
  \item A Secure Channel for communication with
corresponding controller(s).
  \item OpenFlow as the standard protocol for
communication with controller(s) \cite{mckeown2008openflow}.
\end{itemize}

More details on the OpenFlow switch are presented in Section
\ref{sec:switchStraucture}.

\subsection{Alloy Modelling Language}
Alloy is a declarative specification language for modeling complex structures
and behaviors in a system. Alloy is based on first order
logic~\cite{Jackson:2006} and is designed for model checking. An Alloy model is a
relational model which may contain:
\begin{itemize}
\item Signatures: represent the entities in a system.
\item Relations: which relates a signature to another signature.
\item Facts: specify the constraints on the signatures which are assumed to
always hold.
\item Predicates: specify the constraints which can be used to show operations.
\item Functions: are alternative names for expressions that return some result.
\item Assertions: constraints which are checked over the model.
\end{itemize}
%Alloy's tool,
The Alloy Analyzer takes an Alloy model and checks it against its constraints. It
translates the Alloy model into a boolean expression which is analyzed by the SAT
solver embedded in the analyzer. The Alloy Analyzer generates instances of model
invariants satisfying model constraints and then checks specified properties on
these instances. It just returns the models within the user-specified scope,
consisting of a finite number of objects. The result is turned into a graphical
representation of a model instance.

Besides finding satisfying instances, the Alloy Analyzer also
checks assertions. If there is a model within the scope which does not satisfy the
assertions, it will return the model as a counterexample. However if no instance
is found, the assertion may still be not valid in a larger scope.
\subsection{Examples:}
The general syntax for defining a signature in Alloy is as following:\\
\texttt{ sig A {fields}}\\
This line defines signature \texttt{A} with some fields. For
instance Figure \ref{exampleCode:sigSwitch} shows two entities in Alloy which
are defined using keyword \textbf{sig}. Line 1 defines the \texttt{Switch} entity
with a field \texttt{Table}. Signature \texttt{SwState} shows the state of the
network at each time epoch. \texttt{SwSatet} has two relations, \texttt{switch}
and a set of buffered messages (\texttt{BuffMsg}) mapping ports to Messages.

\begin{figure}
{\scriptsize
\begin{Verbatim}[frame=single, numbers=left, numbersep=2pt]
 some sig Switch{tables: some Table}
 sig SwState{sw: one Switch, BuffMsg: set (Port->Message)}
\end{Verbatim}
}
\caption{Example of a signature in Alloy.}
\label{exampleCode:sigSwitch}
\end{figure}
Operations that modify the state of network may be modeled as
\textit{predicates} using the \texttt{\textbf{pred}} keyword. The syntax is:\\
\texttt{pred Name [parameters] {f}}\\
This line defines a predicate, with the given name and (possibly empty)
parameters.
A predicate always produces true or false, so no type is needed.
The result is defined by the formula f, which may reference the parameters.
 Figure~\ref{exampleCode:predRecieve} specifies how the state changes when a new
message is received. s indicates the current state and s' indicates the next
state after receiving a new message. Upon receiving a message the only change is
the relations between switch state and the set of buffered messages. The new
message is added to the buffer, but the other relations stay unchanged.\\
\begin{figure}
{\scriptsize
\begin{Verbatim}[frame=single, numbers=left, numbersep=2pt]
 pred receive(s,s' : SwitchState, m:Message){
	s'.switch = s.switch
	s'.sInBuffMsg = s.sInBuffMsg + m
	s'.sOutBuffMsg = s.sOutBuffMsg
	s'.sTables = s.sTables
}
\end{Verbatim}
}
\caption{Example of a predicate in Alloy.}
\label{exampleCode:predRecieve}
\end{figure}
%\subsection{Examples:}
Facts express constraints on signatures and the relations that must always hold
in the model. The  following syntax shows how to define a fact:\\
\texttt{fact Name {e}}\\
You can name a fact if you wish.
The analyzer will ignore the names. The expression \texttt{e} is a constraint
that the analyzer will assume is always true.
For instance the fact in Figure~\ref{exampleCode:factTable} implies the next table
of every switch table should be in the same switch.
\begin{figure}
{\scriptsize
\begin{Verbatim}[frame=single, numbers=left, numbersep=2pt]
fact {all s:Switch, t: s.tables | t.nxTable in s.tables}
\end{Verbatim}
}
\caption{Example of an Alloy fact}
\label{exampleCode:factTable}
\end{figure}
Using following syntax a function can be defined in Alloy:\\
\texttt{ \textbf{fun} Name \textbf{[}parameters\textbf{]} : type
\textbf{{}e\textbf{}}}\\
This line defines a function, with the given name and (possibly empty)
parameters, and outputting a relation (or a set, or scalar) of the given type
\texttt{type}. The result is defined by the expression e, which may reference
the parameters. For instance Figure \ref{exampleCode:funController} presents an
example of a function in Alloy. This function finds the set of switches' ports
(from set of all input ports \texttt{ports}) connected to a set of controllers
(\texttt{c}).\\
\begin{figure}
{\scriptsize
\begin{Verbatim}[frame=single, numbers=left, numbersep=2pt]
fun findCPort(c: set Controller, ports: set Port): set Port
 {
    (connect.(c.ports)) & ports
 }
\end{Verbatim}
}
\caption{Example of a function in Alloy.}
\label{exampleCode:funController}
\end{figure}
An Alloy assertion can be defined using the following syntax:\\
\texttt{assert Name {f}}\\
This line defines a assertion, with the given name.  Assertions take no
parameters. An assertion always produces true or false, so no type is needed.
The property that is going to be checked  is defined by the formula \texttt{f}.
Figure \ref{exampleCode:assertAcyclic} depicts an example of an assertion in
Alloy. \texttt{Acyclic} asserts there is no loop in the tables' chain.  This
assertion is checked in all models with at most 5 elements of each signature.\\
\begin{figure}
{\scriptsize
\begin{Verbatim}[frame=single, numbers=left, numbersep=2pt]
assert Acyclic{ no t: Table | t in t.^nxTable }
check Acyclic for 5
\end{Verbatim}
}
\caption{Example of an assertion in Alloy.}
\label{exampleCode:assertAcyclic}
\end{figure}
So far Alloy's basics which we are going to use in our code are explained. In
the following sections the Alloy code for modelling the OpenFlow switch is
presented.

\section{Switch Structure}
\label{sec:switchStraucture}
\subsection{OpenFlow Tables}
Any Openflow switch has at least one Flow Table. Each table may have a pointer
to another table as the next table. There is no pointer to the first table, and
this table is called the root table.

A Flow Table consists of some flow entries.
Different components of each flow entries are:
%\begin{comment}
 \begin{itemize}
 \item Match fields: to match against packets. These consist of the ingress port
and packet headers, and optionally metadata specified by a previous table.
\item Counters to update for matching packets.
\item Instructions to apply to matching packets.
 \end{itemize}
%\end{comment}
\begin{comment}
\begin{center}
\begin{table}[h!]\footnotesize
\centering
\begin{tabular}{ | l | c | r |}
  \hline
  Match Fields & Counters & Instructions
  \\ \hline
\end{tabular}
  \caption{Main components of a flow entry in a flow table}
  \label{tab:entry}
\end{table}
\end{center}
\end{comment}
\subsubsection{Match Fields}
In our model of an OpenFlow switch a very simplified version of the match fields is modeled.
For a complete set of match fields that a packet is compared against, see the current OpenFlow switch specification \cite{specv1}.
Currently a packet is compared against the ingress port, source IP and
destination IP.
Each match field in a flow table has a priority field.
If more than one match fields match with an incoming packed, the one with higher
priority will be triggered.
\begin{comment}
\begin{center}
\begin{table}[h!]\scriptsize
\centering
\begin{tabular}{ | l | c | r |  l | c | r |  l | c | r |  l | c | r | l | c | r
|}
  \hline
 \begin{turn}{90}  Ingress Port  \end{turn} & 		
 \begin{turn}{90} Metadata  \end{turn} &
\begin{turn}{90} Ether src \end{turn} &
\begin{turn}{90} Ether dst \end{turn} &
\begin{turn}{90} Ether type \end{turn} &
\begin{turn}{90} VLAN id \end{turn} &
\begin{turn}{90}  VLAN priority  \end{turn} &
\begin{turn}{90} MPLS label  \end{turn} &
\begin{turn}{90} MPLS traffic class \end{turn} &
\begin{turn}{90} IPv4 src \end{turn} &
\begin{turn}{90} IPv4 dst \end{turn} &
\begin{turn}{90} IPv4 proto / ARP opcode IPv4  \end{turn} &
\begin{turn}{90} ToS bits \end{turn} &
\begin{turn}{90} TCP/ UDP / SCTP src port  ICMP Type \end{turn} &
\begin{turn}{90} TCP/ UDP / SCTP dst port
ICMP Code \end{turn}
  \\ \hline
\end{tabular}
  \caption{Fields from packets used to match against flow entries}
  \label{tab:match}
\end{table}
\end{center}
\end{comment}
\subsubsection{Counters}
Counters are stored statistics that can be maintained for each flow, port,
table, \emph{etc.}
In current work, a very simplified version of counters in the static model of
OpenFlow switches is presented.

\subsubsection{Overview}
A set of instructions is associated with every flow entry.
This set of instructions is executed whenever an incoming packet matches the
corresponding entry. In this version of the OpenFlow switch model the following
instructions are modeled:
\begin{itemize}
  \item \textbf{Apply-Action \emph{action(s)}:} A specific set of actions is
immediately applied while the current action set (associated with the packet) remains unchanged.
  \item \textbf{Clear-Action:} clear the associated action set.
  \item \textbf{Write-Action \emph{action(s)}:} Add a specific set of actions to
the current actions set.
  \item \textbf{Goto-Table \emph{next-table-id}:} continue the pipe-lining process
(described in next section) from  table with ID equal to next-table-id. The
next-table-id must be greater than the current table-id.
\end{itemize}
\subsubsection{Action Set}
Each incoming packet has an action set which is initially empty.
Using \emph{Write-Action} and \emph{Clear-Action}, the action set can be
modified whenever the packet matches an entry.
An action set contains at most one action of each type.
In order to have multiple actions of the same type, the \emph{Apply-Action}
instruction can be used.
There are currently two types of action in the OpenFlow specification: required and optional \cite{specv1}.
No optional action is modeled in the current work.
Supported required actions in this model are:\\ \\
\emph{output}:
  \begin{itemize}
    \item \textbf{Forward-to-Port \emph{port}:} forward the packet to
\emph{port}.
    \item \textbf{Forward-to-Controller \emph{controller}:} forward the packet
to a specific controller.
    \item \textbf{Forward-to-Ingress:} forward the packet to the ingress port
(the port that packet has been received from).
    \item \textbf{Forward-to-All:} forward the packet to all outgoing ports.
  \end{itemize}
\emph{Drop}: Drop the packet. This action also can be applied implicitly for
those packet whose action sets have no output action.

\subsection{Matching a packet with flow entries}
\begin{figure}[!]
\begin{center}
\includegraphics[scale=.35]
    {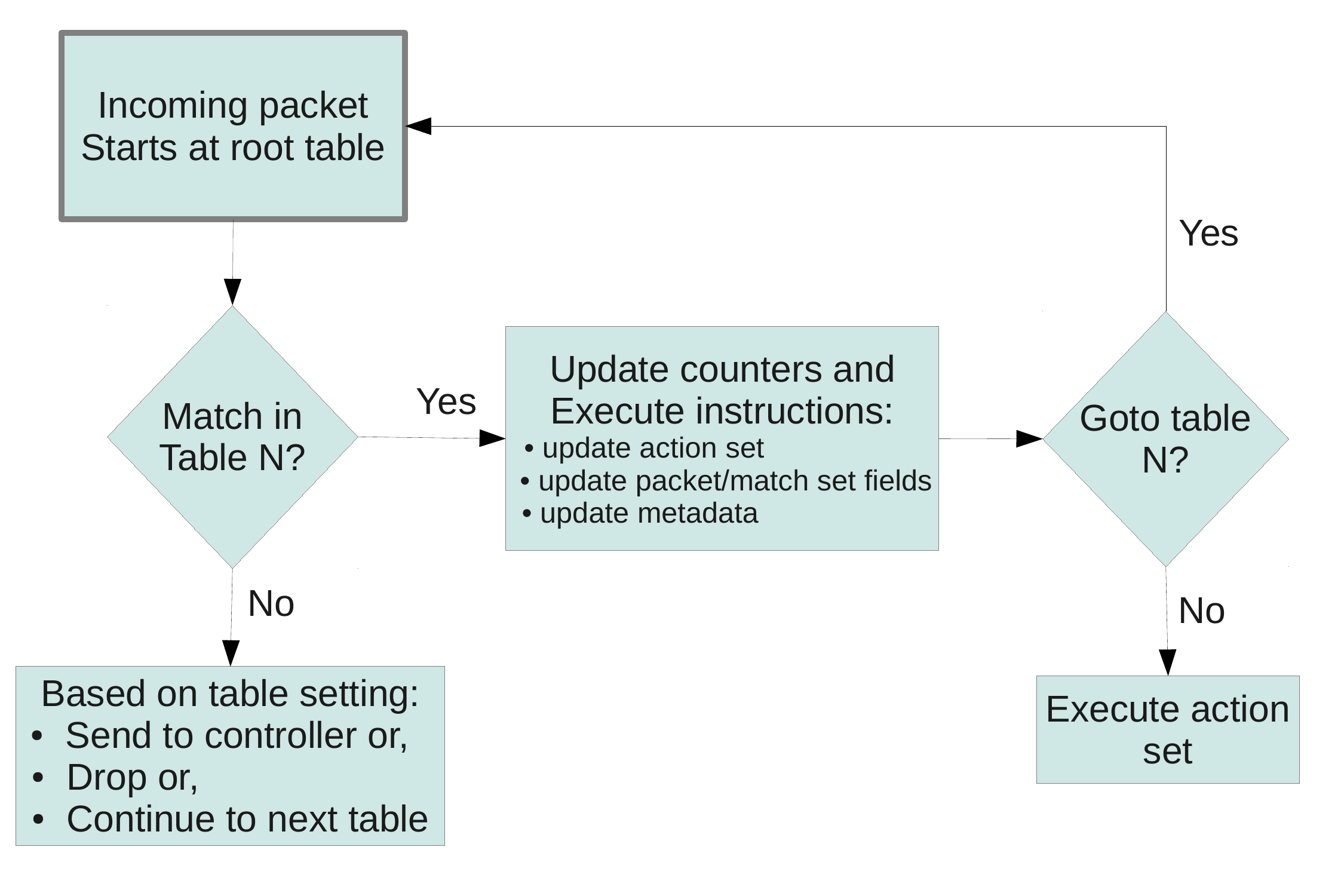}
\caption{Packet flow through an OpenFlow switch.}
\label{fig:PL-flowchart}
\end{center}
\end{figure}

If a packet matches a flow entry in a flow table, the corresponding
instruction set is executed. The instructions in the flow entry may explicitly
direct the packet to another flow table, where the same process is repeated
again. A flow entry can only direct a packet to a flow table number which is
greater than its own flow table number, in other words pipeline processing can
only go forward and not backward. Obviously, the flow entries of the last table
of the pipe-line can not include the Goto instruction. If the matching flow
entry does not direct packets to another flow table, pipe-line processing stops
at this table and the corresponding action set will be executed. Packet flow
through an OpenFlow switch is presented in Figure~\ref{fig:PL-flowchart}.

\section{Static Model of OpenFlow Switch}
\label{sec:static}
Our Alloy model enables us to model states of one switch and its interaction
with other nodes in the network.
We first describe the entities in our model and then we introduce the
constraints on the entities.
These constraints (modeled using Alloy facts) help us to model the correct
structure of an OpenFlow network.

\subsection{Entities}
An OpenFlow network consists of a set of nodes(controller and switches).
Every node has a set of ports which connect nodes together.
In addition to ports, switches also have a set of flow tables.
The root table is the first table which the pipe-lining process for an incoming
packet starts with.
Every packet has a source and a destination IP.
If packet is coming from the controller, it also includes a \texttt{preTable}
and a \texttt{postTable} which reflect the modification need to be done in
switch's flow tables.
Using this modification (\texttt{preTable} will be replaced by
\texttt{postTable}).
In our model each packet is encapsulated in a message which specifies the type
of message and the port the packet was received on.
The Alloy model of different elements of an OpenFlow network is presented in
Figure~\ref{code:sigNodes}.

\begin{figure}
{\scriptsize
\begin{Verbatim}[frame=single, numbers=left, numbersep=2pt]
 abstract sig Node{contained:one Network, ports:some Port}
 some sig Switch extends Node{tables: some Table, root: one
  RootTable}
 some sig Controller extends Node{control:some Switch}
 sig Port{connect:lone Port}
 one sig Network{content:some Node}
 sig Packet{srcIP: one IP, destIP: one IP, preTable,
  postTable :  lone Table }
 some sig Message{msgInPort: one Port, packet: one Packet,
  type: one MsgType}
 abstract sig MsgType{}
 one sig ControllMsgType extends MsgType{}
 one sig HostMsgType extends MsgType{}
\end{Verbatim}
}
\caption{A network model consists of nodes, packets and messages. Lines 12 and
13 model two types of incoming messages.}
\label{code:sigNodes}
\end{figure}
The most important part of the switch model is the flow tables.
The tables are uniquely determined by their tableID and all the tables (except
the last table) have a pointer to the next table.
Every flow table includes a set of flow entries, which messages are checked
against.
If none of the entries match a message, a table miss occurs.
Table miss can be continuing pipe-line process with the next table, forwarding
message to some controllers or dropping the packet.
Tables and related elements are modeled depicted in Figure~\ref{code:sigTable}.

\begin{figure}
{\scriptsize
\begin{Verbatim}[frame=single, numbers=left, numbersep=2pt]
 sig Table {nxTable:lone Table, entries: set FlowEntry,
  tableID: one Int,  miss: lone TableMiss}
 sig RootTable extends Table {}
 abstract sig TableMiss{}
 one sig MissDrop extends TableMiss{}
 one sig MissNext extends TableMiss{}
\end{Verbatim}
}
\caption{Every flow table contains some flow entries and has a corresponding
action for packet miss occurrence.}
\label{code:sigTable}
\end{figure}
As described in section \ref{sec:switchStraucture} each flow entry contains
match fields, counter and instructions.
Our simplified modeling of different types of actions and instructions and their
relation with each other and with respect to the flow entries is presented in
Figure~\ref{code:sigFlowEntry}.
For more detail on instructions and actions and their relation you can refer to
section \ref{sec:switchStraucture} or OpenFlow specification \cite{specv1}.
\begin{figure}
{\scriptsize
\begin{Verbatim}[frame=single, numbers=left, numbersep=2pt]
 sig FlowEntry {match: one MatchField, counters: some
  Counter, instructs: one Instruction, flowID: one Int}
 abstract sig Instruction{}
 sig ApplyActionsIntruct extends Instruction{appActions:
  set Action}
 sig WriteActionsIntruct extends Instruction{wrtActions:
  set Action}
 sig ClearActionIntruct extends Instruction{}
 sig GotoTableIntruct extends Instruction{gotoTable: one
  Table}
 abstract sig Action {}
 sig ForwardtoPort extends Action{toPort: one Port}
 one sig ForwardtoAll extends Action{}
 one sig ForwardtoIngress extends Action{}
 sig ForwardtoController extends Action{toController: one
  Controller}
one sig Drop extends Action{}
\end{Verbatim}
}
\caption{Definition of flow entries, different instructions and actions.}
\label{code:sigFlowEntry}
\end{figure}
\subsection{Constraints on Network Entities}
The model of OpenFlow switch imposes some constraints on the network entities.
These constraints help us to get a correct model of structure of an OpenFlow
network.
In this section we will try to capture those important rules.
The fact in Figure~\ref{code:factSwitch} ensures that all of the switches are
controlled by at least one controller and the controller and switch are
connected.
\begin{figure}
{\scriptsize
\begin{Verbatim}[frame=single, numbers=left, numbersep=2pt]
 fact{all s: Switch |some p1: s.ports, c:Controller, p2:
  c.ports | s in c.control && p2 in p1.connect}
\end{Verbatim}
}
\caption{Every OpenFlow switch is controlled by at least one controller and they
are connected to each other.}
\label{code:factSwitch}
\end{figure}
Based on the OpenFlow specification, each switch has at least one flow table.
There is exactly one table as the root table which the pipe-lining process
starts from.
The set of tables of a switch create a loop free chain.
Each table has some (at least one) flow entry.
Also each table has a table miss action.
This table miss action can be explicit or implicit.
Implicit means there is no specific action assigned for a table, hence the
default action must be applied.
Different table miss actions and the behaviour of a switch for each one is
presented in section \ref{sec:dynamic}.
In order to make sure that our model conforms to all these rules, a group of Alloy
rules are needed.
The combination of these rules presented in Figure~\ref{code_fact:table} ensure a model representative of the specification.
For instance, lines 11-18 take care of following rules: 1) there is only one
connected acyclic chain of tables in each switch,
2) all of the tables in the chain belong to the same switch and 3) there is only
one last table in this chain which its pointer to next table is empty.
%Line 1 makes sure every table belongs to exactly one switch and not more.
%Line 2 and 3 take care of the rule that each root table belongs to exactly one switch and each table miss belongs to one or more tables.
%Line 4 makes sure there is no empty table, there should be at least one entry in each table.
%Line 5 implies that each switch has exactly one root table.
%Line 6 says that root table is a subset of table relations.
%Hence all the rules on the tables hold for root table as well.
%Lines 7, 8 and 9 imply that the root table ID is 0 and all other table IDs
% should be grater than 0, also for all tables  the next tables' ID (obviously if
% exists) should be tables' ID plus one.
% The other constraint in this figure worth mentioning is that line 19 dictates
% that the table miss action of the last table in the chain, cannot be forwarding
% to the next table (simply because there is no next table).
\begin{figure}
{\scriptsize
\begin{Verbatim}[frame=single, numbers=left, numbersep=2pt]
 fact{all t: Table | one s: Switch |  t in s.tables}
 fact{all t: RootTable | one s: Switch |  t in s.root}
 fact {all m:TableMiss | some t:Table | m in t.miss}
 fact {all t:Table | #(t.entries) > 0}
 fact {all s : Switch | one r: RootTable | r in s.tables}
 fact{all s:Switch | s.root in s.tables}	
 fact {all t:RootTable | t.tableID = 0}
 fact {all t:Table | t.tableID >= 0}
 fact nextTableID{all t,t':Table | (t' = t.nxTable) implies
  (t'.tableID = t.tableID.plus[1])}
 fact {all s: Switch, t:s.tables | !(t in  RootTable)
  implies (one t':Table | t'.nxTable = t)}
 fact {all s:Switch, t: s.tables | t.nxTable in s.tables}	
 fact{no t: Table | t.nxTable = t}
 fact {all s:Switch | one t: s.tables | no t.nxTable}
 fact acyclicTable { no t: Table | t in t.^nxTable }
 fact {all s:Switch, t, t': s.tables | !(t = t') implies
  !(t.nxTable = t'.nxTable)}
 fact {all t:Table | #(t.nxTable) = 0 implies !(MissNext
  in t.miss)}
  }
\end{Verbatim}
}
\caption{}
\label{code_fact:table}
\end{figure}
All constraints on flow entries are presented in
Figure~\ref{code:factflowEntry}.
Each flow entry is contained in only one table and every counter is contained in
at least one flow entry.
Line 4 also implies that every action is used in at least one instruction.
\begin{figure}
{\scriptsize
\begin{Verbatim}[frame=single, numbers=left, numbersep=2pt]
 fact {all e: FlowEntry | one t: Table | e in t.entries}
 fact {all c: Counter | some e: FlowEntry | c in
  e.counters}
 fact {all a: Action | (a in ApplyActionsIntruct.appActions)
  or (a in WriteActionsIntruct.wrtActions)}
\end{Verbatim}
}
\caption{All related facts on flow entries of a table.}
\label{code:factflowEntry}
\end{figure}
Obviously in any network model each node (switch or controller) has some ports.
Every port belongs to exactly one node and it cannot be connected to another port
in the same node.
Also each port can be connected to at most one other port and all connections
are two ways.
All these structural rules are applied using facts in
Figure~\ref{code_fact:table}.
In our model we are not interested in relations between controllers, hence line
8 imposes that controller's ports should be connected only to switch's ports.
\begin{figure}
{\scriptsize
\begin{Verbatim}[frame=single, numbers=left, numbersep=2pt]
 fact {all p: Port | one n:Node | p in n.ports}
 fact {no n:Node, p: n.ports | p.connect in n.ports}
 fact {all p: Port | #(p.connect) <=1}
 fact {all p, p': Port | (p = p'.connect) <=>
  (p' = p.connect)}
 fact{all s: Switch |some p1: s.ports, c:Controller, p2:
  c.ports | s in c.control && p2 in p1.connect}
  fact {all p: Port | p in Controller.ports implies
   p.connect in Switch.ports}
\end{Verbatim}
}
\caption{Constraints on the ports and connections between them.}
\label{code_fact:port}
\end{figure}
The first fact in Figure~\ref{code_fact:instrcution} implies every instruction
should belong to at least one flow entry.
In addition in the \texttt{GotoTableIntruct} instruction, the pointed table must
be in the corresponding switch and it must be a table with larger ID.
\begin{figure}
{\scriptsize
\begin{Verbatim}[frame=single, numbers=left, numbersep=2pt]
 fact {all i: Instruction | some f: FlowEntry | i in
  f.instructs}
 fact {all s:Switch, t:s.tables, e: t.entries, i:
  e.instructs | (i in GotoTableIntruct) implies
  ((i.gotoTable in s.tables) and (i.gotoTable.tableID >
  t.tableID) )}
\end{Verbatim}
}
\caption{Facts on flow entry's instructions.}
\label{code_fact:instrcution}
\end{figure}
In Figure~\ref{code_fact:action} the set of facts on different type of actions
is given.
The second fact in line 4 imposes that the \texttt{toPort} of an
\texttt{ForwardtoPort} action must be in the same switch.
Finally lines 8 and 10 make sure there is no repeated actions in our model.
\begin{figure}
{\scriptsize
\begin{Verbatim}[frame=single, numbers=left, numbersep=2pt]
 fact {all a: Action | (a in
  ApplyActionsIntruct.appActions) or (a in
  WriteActionsIntruct.wrtActions)}
 fact {all a: ForwardtoPort, s:Switch | ( (a in
  s.tables.entries.instructs.appActions) or  (a
  in s.tables.entries.instructs.wrtActions) ) implies
  (a.toPort in s.ports) }
 fact {all f, f':ForwardtoPort | !(f = f') implies
  !(f.toPort = f'.toPort) }
 fact {all f, f':ForwardtoController | !(f = f') implies
  !(f.toController = f'.toController) }
\end{Verbatim}
}
\caption{Facts on actions in a flow table.}
\label{code_fact:action}
\end{figure}

\section{Dynamic Behavior of the Model}
\label{sec:dynamic}
\begin{figure} [!]
\begin{center}
\includegraphics[scale=.4]
    {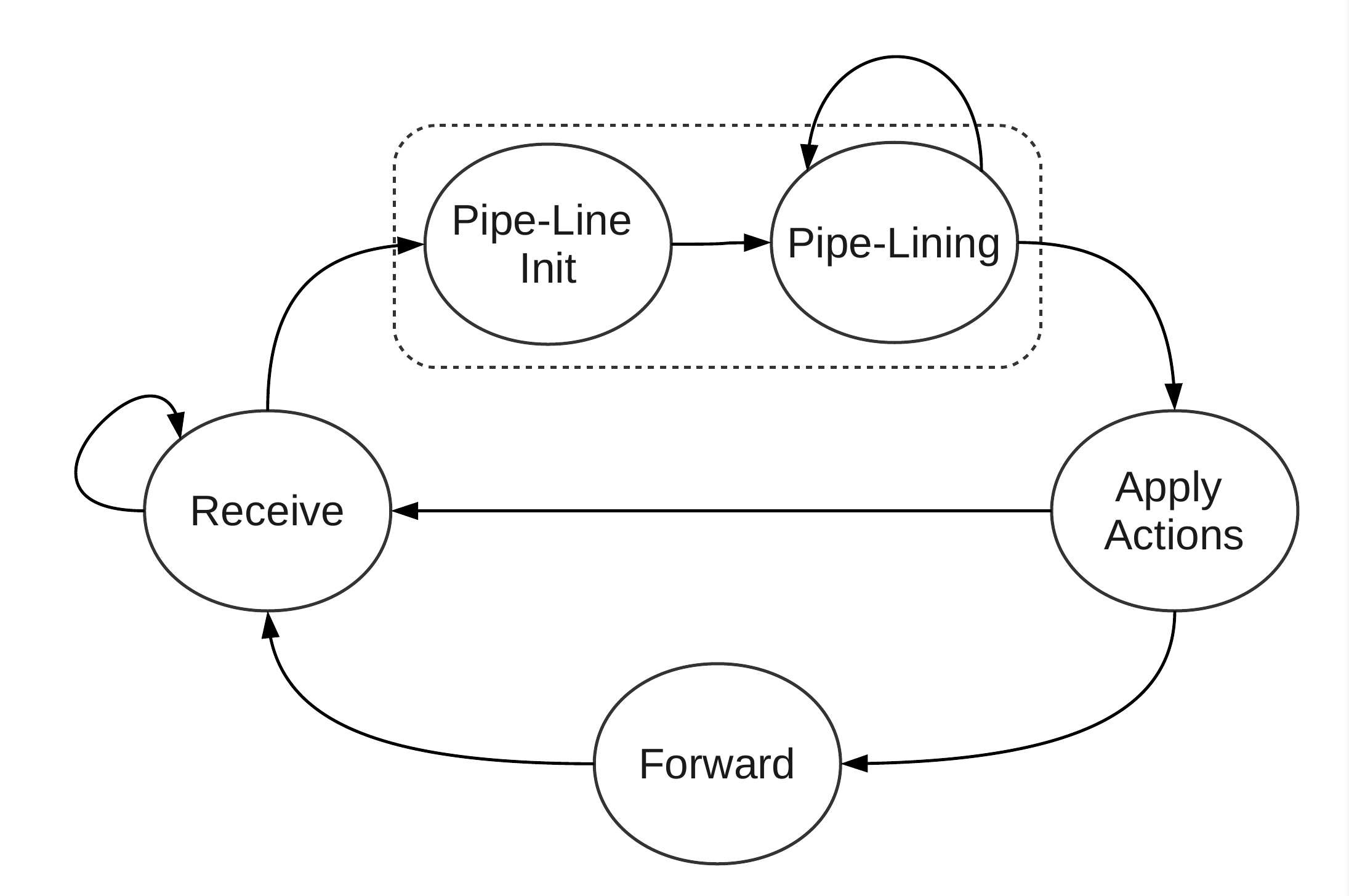}
\end{center}
\caption{Five possible state transitions that are considered in our modeling.}
\label{fig:stateTransition}
\end{figure}

In the previous section the static model of an OpenFlow switch network was presented.
Assuming that this model presents the (simplified) real structure of a OpenFlow
switch network correctly, the dynamic behavior of OpenFlow switch can be modeled
now.
Without loss of generality, the internal behavior of only one switch in relation
with other elements of network is considered.
The internal state of a specific switch changes by various events such as
receiving, sending or processing of a message and \textit{etc}.
In Figure~\ref{code:sigSwitchState} you can see how the state of an OpenFlow
switch is captured.
As you see every element that may change in a switch is considered as a part of
the internal state of a switch.
\begin{figure}[!]
{\scriptsize
\begin{Verbatim}[frame=single, numbers=left, numbersep=2pt]
 sig SwitchState{
  switch: one Switch, sTables: some Table,
  sInBuffMsg: set  Message,
  sOutBuffMsg: set(Port-> Message),
  sTEntries: set FlowEntry, pInHistory: set Message,
  pOutHistory: set (Port->Message),
  nxtInPLTable: lone Table,
  inPL: lone Message, actionSet: set Action,
  outPL: lone Message
 }
\end{Verbatim}
}
\caption{Definition of a switch's state.
Important elements of a \texttt{SwitchState} are \texttt{sInBuffMsg},
\texttt{sOutBuffMsg}, \texttt{pInHistory} and \texttt{pOutHistory}.
Respectively these elements present the set of buffered for processing, buffered
for forwarding, saved in input history and saved in output history messages.
Beside, \texttt{nxtInPLTable}, \texttt{inPL}, \texttt{actionSet} and
\texttt{outPL} are used to handle the behavior of switch during the switch's
table pipe-lining process.
\texttt{nxtInPLTable} keeps the next table that must be used for flow entry
matching.
\texttt{inPL} and \texttt{outPL} denote the message that is checked against flow
entries in pipe-lining process.
\texttt{actionSet} keeps a record of actions that are being added whenever a match
is found for the message.
Also notice that since the by arrival of a \textit{control Messages}, the flow
entries of the tables may change.
Hence the set of tables and entries of a switch is also kept in the switch's
state (modeled by \texttt{sTables} and \texttt{sTEntries} respectively).}
\label{code:sigSwitchState}
\end{figure}
%%%%%%%%%%%%%%%%
In order to have a correct chain of switch states, the first
\texttt{SwitchState} must be initialized. For instance the set of input/output
buffered messages in this state must be empty. This rule is applied to the first
\texttt{SwitchState} using Alloy fact presented in Figure~\ref{code:factFirst}.
%%%%%%%%%%%%%%%
\begin{figure} [!]
\begin{center}
{\scriptsize
\begin{Verbatim}[frame=single, numbers=left, numbersep=2pt]
 fact {
  no first.sInBuffMsg && no first.pInHistory &&
  no first.pOutHistory && no first.sOutBuffMsg  &&
  first.switch.tables = first.sTables  &&
  no first.sTEntries && no first.nxtInPLTable &&
  no first.inPL && no first.actionSet &&
  no first.outPL
 }
\end{Verbatim}
}
\caption{Using this fact the first state is initialized. \texttt{first} is a
reserved keyword in \texttt{order} library referring to the first
\texttt{SwitchState}.}
\label{code:factFirst}
\end{center}
\end{figure}
The core part of modeling the dynamic behavior of the switch is to capture the
possible transitions between different states.
Possible transitions that we have considered in our model are schematically
presented in Figure~\ref{fig:stateTransition}.
In our model the pipe-lining procedure is considered as an atomic transition.
Hence until the whole pipe-lining is done, no new message will be received
or forwarded.
In order to make modeling easier the pipe-lining is broken down into two parts,
\textit{Pipe-Line initiation} and \textit{Pipe-Lining}.
All these five state transitions are implemented and described by five major
Alloy predicates as we explain in following subsections.

\subsection{Receiving a Message}
The first state change happens by receiving a new message.
Hence the only change in switch state is adding a new message into the input buffer
and input history of the switch.
All other elements of \texttt{SwitchState} are remained unchanged by receiving a
new message.
These rules are applied by two conditions: \texttt{ss'.sInBuffMsg =
ss.sInBuffMsg + m} and \texttt{ss'.pInHistory = ss.pInHistory + m}.
This transition is modeled using the receive predicate presented in
Figure~\ref{code:pred-recieve}.
\begin{figure} [!]
{\scriptsize
\begin{Verbatim}[frame=single, numbers=left, numbersep=2pt]
 pred recieve(ss,ss' : SwitchState, m:Message){
  #(Port->m & (ss'.pOutHistory)) = 0
  #(Port->m & (ss.pOutHistory)) = 0
  #(Port->m & (ss'.sOutBuffMsg)) = 0
  #(Port->m & (ss.sOutBuffMsg)) = 0
  !(m in ss.sInBuffMsg)
  ss'.sInBuffMsg = ss.sInBuffMsg + m
  !(m in (ss.pInHistory))
  ss'.pInHistory = ss.pInHistory + m
  ss'.sInBuffMsg in ss'.pInHistory
  ss.sInBuffMsg in ss'.pInHistory
  ss'.pOutHistory = ss.pOutHistory
  ss'.sOutBuffMsg = ss.sOutBuffMsg
  ss'.sTables = ss.sTables
  ss'.sTEntries = ss.sTEntries
  no(ss'.actionSet)
  no(ss.actionSet)
  no(ss'.outPL)
  no(ss.outPL)
  no(ss'.nxtInPLTable)
  no(ss.nxtInPLTable)
  no(ss'.inPL)
  no(ss.inPL)
  ss'.switch = ss.switch
}
\end{Verbatim}
}
\caption{Predicate receive: $\texttt{ss}$ represent the state of switch before
receiving the message $\texttt{m}$ and $\texttt{ss'}$ is the state of switch
after receiving message $\texttt{m}$. The first four lines apply the condition
that the new message $\texttt{m}$ is not a repeated message.
In line 7 and 9 it's stated that the only difference between $\texttt{ss}$ and
$\texttt{ss'}$ is addition of message $\texttt{m}$ to input buffer and input
history.
Other elements of the switch's state are conditioned to remain unchanged in this
predicate.}
\label{code:pred-recieve}
\end{figure}

\subsection{Pipe-Line Initiation}
As mentioned before a received message is buffered in the switch and recorded in
the input history. In general, a message is either a control message or a host message.
Control messages are used to change the flow entries or flow tables in the
switch. If the message is a host message, however, pipelining can be initiated.
Predicate \texttt{pipeLineInit} as presented in Figure~\ref{code:pred-PLInit}
shows this state transition.
In this predicate condition $\texttt{ss'.sInBuffMsg = ss.sInBuffMsg - m}$ makes
sure that message $\texttt{m}$ is removed from the input buffer.
Lines 17 to 27 handle the message based on its type.
Lines 17 to 22 model the behavior of switch for a control message.
In these lines the set of tables of switch (modeled by \texttt{sTables}) changes
by replacing the \texttt{preTable} with \texttt{postTable} of arrived message
from controller.
The way that flow entries of tables are manipulated in our model is slightly
different from the OpenFlow specification~\cite{specv1}.
In our model we have simplified this process.
Instead of removing or adding a single entry in a specific table, the table is
replaced with the updated table.
In lines 24 to 27 if the message is of type host message it is set to start
pipe-lining in the next state.
\begin{figure} [!]
{\scriptsize
\begin{Verbatim}[frame=single, numbers=left, numbersep=2pt]
 pred pipeLineInit(ss,ss' : SwitchState, m:Message){
  ss'.switch = ss.switch
  ss'.sInBuffMsg = ss.sInBuffMsg - m
  ss'.sOutBuffMsg = ss.sOutBuffMsg
  ss'.pInHistory = ss.pInHistory
  (m in ss.sInBuffMsg)
  (m in ss.pInHistory)
  #(Port->m & (ss.pOutHistory)) = 0
  #(Port->m & (ss.sOutBuffMsg)) = 0
  ss'.pOutHistory = ss.pOutHistory
  no(ss.nxtInPLTable)
  no(ss.inPL)
  no(ss.outPL)
  no(ss.actionSet)
  no(ss'.actionSet)

  m.type = ControllMsgType implies
   ((ss'.sTables = ss.sTables - m.packet.preTable
   + m.packet.postTable) and (ss'.sTEntries = ss.sTEntries
   - m.packet.preTable.entries + m.packet.postTable.entries)
   and (no ss'.nxtInPLTable) and (no ss'.inPL) and
   (no ss'.outPL))

  m.type = HostMsgType implies
   (ss'.sTables = ss.sTables and ss'.inPL = m and
    ss'.outPL = m and ss'.nxtInPLTable = ss.switch.root and
    ss'.sTEntries = ss.sTEntries)
}
\end{Verbatim}
}
\caption{Predicate $\texttt{pipeLineInit}$: Same as predicate
$\texttt{recieve}$, $\texttt{ss}$ represent the state of switch before happening
of pipe-line initiation and $\texttt{ss'}$ is the state of switch after
initializing message $\texttt{m}$ in pipe-line.
In lines 24 to 27 if the message is of type host message, using condition
$\texttt{ss'.inPL = m and ss'.outPL = m}$ message it is set to start
pipe-lining. Also the condition $\texttt{ss'.nxtInPLTable = ss.switch.root}$
implies that the pipe-lining must starts with the root table of the switch.}
\label{code:pred-PLInit}
\end{figure}

\subsection{Pipe-Lining Predicate}
The core of the modeling of the dynamic behavior of a
OpenFlow switch is the pipe-lining process.
In this procedure the message is pulled out of the input buffer and, starting from the root
table, is compared with flow entries.
This predicate has two important parts.
In every round of pipe-lining either there is exactly one flow entry in the
current table in pipe-line that matches the message, or no entry matches it.
In the former case using predicate \texttt{applyInstruction} (presented in
Figure~\ref{code:pred-appInstruct}) the set of instructions for the matched flow
entry is applied to the message.
On the other hand, in the latter case, we are facing a table miss.
Hence the appropriate action (based on the table's miss action) must be carried out.
the table's miss action can be dropping the message, carrying pipe-lining up using
subsequent table or the default action, which is sending the message to connected
controller(s).

The pipe-lining procedure is modeled using predicate \texttt{pipeLining}
presented in Figure~\ref{code:pred-PLing}.
\begin{figure} [!]
{\scriptsize
\begin{Verbatim}[frame=single, numbers=left, numbersep=2pt]
 pred pipeLining(ss,ss' : SwitchState){
  ss'.switch = ss.switch
  ss'.sTables = ss.sTables
  ss'.sInBuffMsg = ss.sInBuffMsg
  ss'.sTEntries = ss.sTEntries
  ss'.pInHistory = ss.pInHistory
  ss'.pOutHistory = ss.pOutHistory
  #(ss.inPL) = 1

  (
   one f:ss.nxtInPLTable.entries |
    match[ss, ss.inPL, f, ss.nxtInPLTable] and
    applyInstruction[ss, ss', f] and
    ss'.sOutBuffMsg = ss.sOutBuffMsg
    and ss'.outPL = ss.outPL)
   		  or
   all f:ss.nxtInPLTable.entries|
    !match[ss, ss.inPL, f,  ss.nxtInPLTable] and
    (
     ((ss.nxtInPLTable.miss in MissDrop) and
      (no ss'.nxtInPLTable) and (no ss'.inPL) and
      (no ss'.actionSet) and
      (ss'.sOutBuffMsg = ss.sOutBuffMsg) and
      (no ss'.outPL))
  			or
     ((ss.nxtInPLTable.miss in MissNext) and
      (ss'.nxtInPLTable = ss.nxtInPLTable.nxTable) and
      (ss'.inPL = ss.inPL) && (ss'.actionSet =ss.actionSet)
      (ss'.sOutBuffMsg = ss.sOutBuffMsg) and
      (ss'.outPL = ss.outPL))
	  		or
     ((no ss.nxtInPLTable.miss) and (no ss'.nxtInPLTable)
      and (no ss'.inPL) and  (no ss'.actionSet) and
      (ss'.sOutBuffMsg = ss.sOutBuffMsg +
      findCntrlPort[Controller, ss.switch.ports]->ss.inPL)
      and (no ss'.outPL))
   )
  )
 }
\end{Verbatim}
}
\caption{Predicate $\texttt{pipeLineInit}$: Same as predicate
$\texttt{recieve}$, $\texttt{ss}$ represent the state of switch before carrying
out pipe-lining on one table and $\texttt{ss'}$ is the state of switch after
pipe-lining. In each round of pipe-lining a message is compared againt the flow
entries of table \texttt{nxtInPLTable}. If a match is found, using predicate
\texttt{applyInstruction} (lines 11 to 15) corresponding instructions are
applied to the packet. If no match is found, then based on table's miss action
one of conditions of lines 17 to 37 must be true.}
\label{code:pred-PLing}
\end{figure}
\begin{figure} [!]
{\scriptsize
\begin{Verbatim}[frame=single, numbers=left, numbersep=2pt]
 pred match(m:Message,f:FlowEntry,t:Table){
  simpleMatch[m,f,t] and
  (no f': FlowEntry |
    !(f = f') and simpleMatch[m, f', t] and
    f'.match.priority < f.match.priority)
 }

 pred simpleMatch(m:Message, f: FlowEntry, t:Table){
  f in t.entries
  f.match.matchPort = m.msgInPort or no f.match.matchPort
  f.match.srcIP = m.packet.srcIP or no f.match.srcIP
  f.match.destIP = m.packet.destIP or no f.match.destIP
  (#(f.match.matchPort)!=0 or
  	#(f.match.srcIP)!=0 or #(f.match.destIP)!=0))
 }
\end{Verbatim}
}
\caption{Predicate $\texttt{match}$: This predicate checks if message
$\texttt{m}$ matches the flow entry $\texttt{f}$ in table $\texttt{t}$ and more
importantly there is no other flow entry in $\texttt{t}$ with lower priority
that matches $\texttt{m}$ (based on matching rule in OpenFlow
specification ~\cite{specv1}). }
\label{code:pred-match}
\end{figure}

\begin{figure} [!]
{\scriptsize
\begin{Verbatim}[frame=single, numbers=left, numbersep=2pt]
 pred applyInstruction(ss,ss' : SwitchState, f: FlowEntry){
  (#(f.instructs & ClearActionIntruct) > 0) implies
   (ss'.actionSet =
     (f.instructs & WriteActionsIntruct).wrtActions)
  else
   (ss'.actionSet = ss.actionSet +
     (f.instructs & WriteActionsIntruct).wrtActions)

  (#(f.instructs & GotoTableIntruct) = 1) implies
   (ss'.nxtInPLTable =
     (f.instructs & GotoTableIntruct).gotoTable and
     ss'.inPL = ss.inPL)
  else
   ((no ss'.inPL) and (no ss'.nxtInPLTable)	)
}
\end{Verbatim}
}
\caption{Predicate $\texttt{applyInstruction}$: This predicate applies the set
of instructions of flow entry \texttt{f} to the message that is currently in
pipe-line process.}
\label{code:pred-appInstruct}
\end{figure}

\subsection{Message Forwarding Predicate}
After being pipelined, a message is usually forwarded via a switch port.
Hence a message may be added to some ports' output buffer.
\texttt{forward} predicate (Figure~\ref{code:pred-forward} ) models the process
of removing one message from output buffer, and forwarding it via the
corresponding port.
\begin{figure} [!]
{\scriptsize
\begin{Verbatim}[frame=single, numbers=left, numbersep=2pt]
 pred forward(ss,ss' : SwitchState, m:Message, p:Port){
  ss'.switch = ss.switch
  ss'.sTables = ss.sTables
  ss'.sTEntries = ss.sTEntries
  (p in ss.switch.ports)
  (p in ss'.switch.ports)
  (p->m in ss.sOutBuffMsg)
  !(m in ss.sInBuffMsg)
  (m in (ss.pInHistory))
  !(p->m in (ss.pOutHistory))
  ss'.sOutBuffMsg = ss.sOutBuffMsg - p->m
  ss'.pOutHistory = ss.pOutHistory + p->m
  ss'.pInHistory = ss.pInHistory
  ss'.sInBuffMsg = ss.sInBuffMsg
  no ss'.nxtInPLTable
  no ss.nxtInPLTable
  no ss'.inPL
  no ss.inPL
  no ss'.actionSet
  no ss.actionSet
  no ss'.outPL
  no ss.outPL
 }
\end{Verbatim}
}
\caption{Predicate $\texttt{forward}$: $\texttt{ss}$ represent the state of
switch before forwarding message $\texttt{m}$ and $\texttt{ss'}$ is the state of
switch after forwarding it via port $\texttt{p}$. As you see the only changes
between $\texttt{ss}$ and $\texttt{ss'}$ are modeled conditions in lines 11 and
12. Respectively in these lines message is removed from output buffer and added
to output history.}
\label{code:pred-forward}
\end{figure}
\subsection{Transition Between States}
In our model it is assumed that transitions are atomic actions.
Namely from any current switch's state \texttt{ss} to the exact next state
\texttt{ss'}, only one of aforementioned five transitions can happen.
In order to apply this rule, we add some conditions in a fact shown in
Figure~\ref{code:fact-transition}.
In this fact it is applied that between any two consecutive state \texttt{ss}
and \texttt{ss'}, exactly one of the discussed predicates can be correct.
In addition each transition happens only for one message.
\begin{figure} [!]
{\scriptsize
\begin{Verbatim}[frame=single, numbers=left, numbersep=2pt]
 fact switchStateTransition{
  all ss: SwitchState, ss' : ss.next {
   (
    (one m:Message | recieve[ss, ss', m] ) and
    (no m:Message, p:Port | forward[ss, ss', m, p] ) and
    (no m:Message | pipeLineInit[ss, ss', m] ) and
    !pipeLining[ss,ss'] and
    !applyActionSet[ss,ss']
   )
	or
   (
    (no m:Message | recieve[ss, ss', m] ) and
    (one m:Message, p:Port | forward[ss, ss', m, p] ) and
    (no m:Message | pipeLineInit[ss, ss', m] ) and
    !pipeLining[ss,ss'] and
    !applyActionSet[ss,ss']
   )
	or
   (
    (no m:Message | recieve[ss, ss', m] ) and
    (no m:Message, p:Port | forward[ss, ss', m, p] ) and
    (one m:Message | pipeLineInit[ss, ss', m] ) and
    !pipeLining[ss,ss'] and
    !applyActionSet[ss,ss']
   )
	or
   (
    (no m:Message | recieve[ss, ss', m] ) and
    (no m:Message, p:Port | forward[ss, ss', m, p] ) and
    (no m:Message | pipeLineInit[ss, ss', m] ) and
    pipeLining[ss,ss'] and
    !applyActionSet[ss,ss']
   )
	or
   (
    (no m:Message | recieve[ss, ss', m] ) and
    (no m:Message, p:Port | forward[ss, ss', m, p] ) and
    (no m:Message | pipeLineInit[ss, ss', m] ) and
    !pipeLining[ss,ss'] and
     applyActionSet[ss,ss']
   )     	
  }
}
\end{Verbatim}
}
\caption{Predicate $\texttt{switchStateTransition}$: between any two switch's
state $\texttt{ss}$ and $\texttt{ss'}$ only one transition happens. Therefore
only one predicate is logically true.}
\label{code:fact-transition}
\end{figure}

\section{Related Work}
\label{sec:related}
In Veriflow \cite{Veriflow:HOTSDN2012} the problem of checking invariants in
software defined networks' data plane in real time is addressed. The authors
proposed to divide the network into equivalent classes so that checking invariants
and violations become easier and more efficient. Conflicting rules can be
detected in real time but rules that rewrite packets cannot be checked by
Veriflow. Anteater \cite{Anteater:Sigcomm2011} statically analyzes the dataplane
configuration to check isolation errors and lack of connectivity due to
misconfigurations. Anteater translates high level network invariants into SAT
instances and use a SAT solver to check them and returns the counterexamples.
Anteater cannot scale well to dynamic network changes and it takes too long to
check invariants.
In \cite{HeaderSpaceAnalysis:NSDI2012} network reachability is examined. To
achieve this, a minimal set of packets which are required to cover all rules or
links in the network is computed. Then in \cite{AutoPacketGeneration:Conext2012}
these packets are sent periodically to all the nodes to check for network
failures and errors.
\cite{WherIsDebuggerForSDN:HOTSDN2012} introduces ndb, a prototype network
debugger inspired by gdb, which implements breakpoints and packet backtraces for
SDN which enables debuggers to track down the cause of an error.
In \cite{NetworkUpdateAbtraction:Sigcomm2012} introduces consistent network
updates in which behaviors are guaranteed to be preserved when a packet it
traversing the network. It will enable us to check consistency after
transitioning between configurations, but it requires us to store a huge number
of extra rules in switches.

%model checking and formulate forwarding loop condition as a temporal logic
%formula. In addition to loop existence, it can checked whether a packet passes
%through a specific switch or not.
\cite{SOFTTestOpenflow:CoNEXT2012} introduces SOFT, an approach to test
interoperability of Openflow switches and inconsistency between different
Openflow agents and the cause of inconsistency.

FlowChecker \cite{FlowChecker:SafeConfig2010} tries to find intra-switch
misconfigurations. FlowChecker translates FlowTable configurations into boolean
expressions using Binary Decision diagrams(BDD) and then checks network
invariants by using model checkers.

In \cite{NetworkConfigBox:ICNP2009} the whole network is modelled as a finite
state machine in which packet header and the location determines states. The
goal of paper is to check correctness of network reachability per packet.To
achieve this, authors have used BDDs and model checking on properties specified
in computation tree logic(CTL) to test all future and past states of packet in
the network .

\cite{ConsistentUpdateforSDN:HotNets2011} studies inconsistencies caused by
updating switch configurations. Authors tries to keep Openflow network
consistent at the cost of
 increasing state in switches to store duplicate table entries.

In addition to switch behavior, a considerable
amount of literature has been published with a focus on Openflow controllers.
For example, NICE \cite{NICE:NSDI2012} is a controller verification tool which
uses model checking and symbolic execution to automate testing Openflow
applications. Nice attempts to explore the state space of the whole network and
find the invalid system states.

\section{Conclusions and Future Work}
\label{sec:conclusionFuture}
The final goal of this research is to use the Alloy Analyzer to generate a model for
lightweight verification of OpenFlow switches in order to help researchers in
analyzing OpenFlow switch networks' properties. So far the modeling of static
structure and also modeling as well as the internal states and dynamic behavior of OpenFlow
switch have been considered, specifically matching rules and actions in table entries as
a part of a real size network. Consequently, there are some
desired properties that we aim to investigate in future. Some important properties that
will be investigated are:
\begin{itemize}
\item Conflicting rules in aggregate flow table
\item Existence of loops in the set of forwarding actions
\item  Scheduling problem in the existence of different controllers
\end{itemize}
In the follow up research, using Alloy Analyzer, we will try to check these
properties of OpenFlow switch networks in our model.

\bibliographystyle{plain} % {alpha} % {siam} % {abbrv} % {plain} %
   % plain, abbrv, siam, alpha, are among dozens of other available styles
\bibliography{allBibs}

\begin{thebibliography}{10}

\bibitem{beacon}
Beacon: a java-based openflow control platform.

\bibitem{NetworkConfigBox:ICNP2009}
E.~Al-Shaer, W.~Marrero, A.~El-Atawy, and K.~Elbadawi.
\newblock Network configuration in a box: towards end-to-end verification of
  network reachability and security.
\newblock In {\em Network Protocols, 2009. ICNP 2009. 17th IEEE International
  Conference on}, pages 123--132, 2009.

\bibitem{FlowChecker:SafeConfig2010}
Ehab Al-Shaer and Saeed Al-Haj.
\newblock Flowchecker: configuration analysis and verification of federated
  openflow infrastructures.
\newblock In {\em Proceedings of the 3rd ACM workshop on Assurable and usable
  security configuration}, SafeConfig '10, pages 37--44, New York, NY, USA,
  2010. ACM.

\bibitem{NICE:NSDI2012}
Marco Canini, Daniele Venzano, Peter Pere\v{s}\'{\i}ni, Dejan Kosti\'{c}, and
  Jennifer Rexford.
\newblock A nice way to test openflow applications.
\newblock In {\em Proceedings of the 9th USENIX conference on Networked Systems
  Design and Implementation}, NSDI'12, pages 10--10, Berkeley, CA, USA, 2012.
  USENIX Association.

\bibitem{NOX:SIGCOMM2008}
{Gude, Natasha and Koponen $et$ $al$}.

\bibitem{WherIsDebuggerForSDN:HOTSDN2012}
Nikhil Handigol, Brandon Heller, Vimalkumar Jeyakumar, David Mazi{\'e}res, and
  Nick McKeown.
\newblock Where is the debugger for my software-defined network?
\newblock In {\em Proceedings of the first workshop on Hot topics in software
  defined networks}, HotSDN '12, pages 55--60, New York, NY, USA, 2012. ACM.

\bibitem{Jackson:2006}
Daniel Jackson.
\newblock {\em Software Abstractions: Logic, Language, and Analysis}.
\newblock The MIT Press, 2006.

\bibitem{HeaderSpaceAnalysis:NSDI2012}
Peyman Kazemian, George Varghese, and Nick McKeown.
\newblock Header space analysis: static checking for networks.
\newblock In {\em Proceedings of the 9th USENIX conference on Networked Systems
  Design and Implementation}, NSDI'12, pages 9--9, Berkeley, CA, USA, 2012.
  USENIX Association.

\bibitem{Veriflow:HOTSDN2012}
Ahmed Khurshid, Wenxuan Zhou, Matthew Caesar, and P.~Brighten Godfrey.
\newblock Veriflow: verifying network-wide invariants in real time.
\newblock In {\em Proceedings of the first workshop on Hot topics in software
  defined networks}, HotSDN '12, pages 49--54, New York, NY, USA, 2012. ACM.

\bibitem{ONIX:2010}
{Koponen, Teemu $et$ $al$}.
\newblock Onix: a distributed control platform for large-scale production
  networks.
\newblock In {\em Proceedings of the 9th USENIX conference on Operating systems
  design and implementation}, OSDI'10, pages 1--6, Berkeley, CA, USA, 2010.
  USENIX Association.

\bibitem{SOFTTestOpenflow:CoNEXT2012}
{Kuzniar, Maciej $et$ $al$}.
\newblock A soft way for openflow switch interoperability testing.
\newblock In {\em Proceedings of the 8th international conference on Emerging
  networking experiments and technologies}, CoNEXT '12, pages 265--276, New
  York, NY, USA, 2012. ACM.

\bibitem{Anteater:Sigcomm2011}
{Mai, Haohui $et$ $al$}.
\newblock Debugging the data plane with anteater.
\newblock In {\em Proceedings of the ACM SIGCOMM 2011 conference}, SIGCOMM '11,
  pages 290--301, New York, NY, USA, 2011. ACM.

\bibitem{POXController}
J~Mccauley.
\newblock Pox: A python-based openflow controller.

\bibitem{mckeown2008openflow}
{McKeown, Nick $et$ $al$}.
\newblock Openflow: enabling innovation in campus networks.
\newblock {\em ACM SIGCOMM Computer Communication Review}, 38(2):69--74, 2008.

\bibitem{NetworkUpdateAbtraction:Sigcomm2012}
Mark Reitblatt, Nate Foster, Jennifer Rexford, Cole Schlesinger, and David
  Walker.
\newblock Abstractions for network update.
\newblock {\em SIGCOMM Comput. Commun. Rev.}, 42(4):323--334, August 2012.

\bibitem{ConsistentUpdateforSDN:HotNets2011}
Mark Reitblatt, Nate Foster, Jennifer Rexford, and David Walker.
\newblock Consistent updates for software-defined networks: change you can
  believe in!
\newblock In {\em Proceedings of the 10th ACM Workshop on Hot Topics in
  Networks}, HotNets-X, pages 7:1--7:6, New York, NY, USA, 2011. ACM.

\bibitem{specv1}
OpenFlow Specification.
\newblock v1. 1.0.

\bibitem{AutoPacketGeneration:Conext2012}
Hongyi Zeng, Peyman Kazemian, George Varghese, and Nick McKeown.
\newblock Automatic test packet generation.
\newblock In {\em Proceedings of the 8th international conference on Emerging
  networking experiments and technologies}, CoNEXT '12, pages 241--252, New
  York, NY, USA, 2012. ACM.

\end{thebibliography}

\end{document}